\documentclass[sigconf,nonacm]{acmart}

\usepackage{algorithm}
\usepackage{algpseudocode}
\usepackage{enumitem}
\usepackage{circledsteps}
\usepackage{siunitx}
\usepackage{csquotes}
\usepackage{booktabs}
\usepackage[noabbrev]{cleveref}

\newcommand{\RNum}[1]{\uppercase\expandafter{\romannumeral #1\relax}}
\begin{document}

\title[Ethereum Needs Fairness Mechanisms that Do Not Depend on  Participants' Altruism]{\texorpdfstring{Why Ethereum Needs Fairness Mechanisms that\\ Do Not Depend on Participants' Altruism}{Why Ethereum Needs Fairness Mechanisms that Do Not Depend on Participants' Altruism}}

\author{Patrick Spiesberger}
\email{patrick.spiesberger@kit.edu}
\orcid{0009-0006-4746-8868}
\affiliation{%
  \institution{Karlsruhe Institute of Technology}
  \city{Karlsruhe}
  \country{Germany}
}

\author{Nils Henrik Beyer}
\email{nils.beyer@student.kit.edu}
\affiliation{%
  \institution{Karlsruhe Institute of Technology}
  \city{Karlsruhe}
  \country{Germany}
}

\author{Hannes Hartenstein}
\email{hannes.hartenstein@kit.edu}
\orcid{0000-0003-3441-3180}
\affiliation{%
  \institution{Karlsruhe Institute of Technology}
  \city{Karlsruhe}
  \country{Germany}
}

\begin{abstract}
Ethereum's ideal of censorship resistance, together with related fairness properties, is undermined in practice, motivating fairness mechanisms that aim to restore these properties.
Several of these mechanisms hand control over block contents to a committee of proposers under a 1-of-\textit{n} honest assumption: at least one committee member complies with the mechanism even when deviating would increase personal revenue.
We refer to such proposers as altruistic.
Yet prior work shows that roughly 91\,\% of blocks are constructed by centralized block-building services that demonstrably take user-adverse actions for financial gain; the responsible proposers sign these blocks blindly, without any means of intervention.
A common reading of this figure is that 9\,\% of proposers forgo these gains and act altruistically.
Our empirical analysis of the full year 2025 shows that this share is far smaller: at most 1.55\,\% of proposers can plausibly be regarded as altruistic, whereas the remaining 98.45\,\% of proposers exhibit observable non-altruistic behavior.
We interpret 1.55\,\% as an upper bound on the prevalence of altruistic proposers.
These results imply that committee-based fairness mechanisms that rely on altruistic members would require substantially larger committees than currently proposed.
This raises concerns about their practical viability and motivates mechanisms in which fair behavior is the rational choice.
\end{abstract}

\keywords{Blockchain, Ethereum, Proposer-Builder Separation (PBS), Maximum Extractable Value (MEV), Fairness, Altruism, Data collection}

\maketitle

\section{Introduction}
\label{intro}
Ethereum aims to be a decentralized, censorship-resistant execution environment~\cite{ethereum_vision}.
In practice, however, Ethereum's block production mechanism~\cite{ETHspec2} grants each proposer\footnote{Every proposer is a randomly selected validator. Anyone can become a validator~\cite{ETHspec2}, but the influence over proposer selection is minimal~\cite{alpturer_optimal_2024}.} temporary monopoly control~\cite{MEV_definition_Sedlmeir} over transaction inclusion.
This enables transaction censorship~\cite{censorship} and user-adverse strategies such as front-running~\cite{MEV_definition_Sedlmeir}.
Empirical evidence~\cite{heimbach_ethereums_2023} shows these behaviors are not only possible but prevalent, and increasingly attractive under stronger economic incentives.
Other work~\cite{yang_decentralization_2025} further shows that these practices have contributed to the centralization of Ethereum's application layer, allowing a small number of powerful actors to maintain control over block contents for extended periods rather than for a single block only.
This raises the question of whether a proposer engages in such practices to maximize profit at the expense of Ethereum's fundamental principles, or instead adheres to the protocol's intended design, with the latter being more conducive to fairness mechanisms.
In cooperative systems such as Ethereum, proposer behavior can be described by the Byzantine--Altruistic--Rational (BAR) model~\cite{aiyer_bar_2005}.
\textit{Byzantine} proposers deviate arbitrarily from the protocol; \textit{rational} proposers deviate whenever doing so increases their own revenue; and \textit{altruistic} proposers follow the protocol without deviation, even when doing so reduces their revenue.
A deviation is not necessarily a protocol violation but rather reflects the exploitation of protocol-granted freedoms by proposers, potentially at the expense of Ethereum's fundamental ideals.
In this paper, we treat altruism as a behavioral notion: altruistic proposers prioritize user fairness and network decentralization over short-term revenue.
Given the large number of proposers and the heterogeneous set of stakeholders operating them, it is unreasonable to assume that the network consists exclusively of rational or Byzantine proposers.
Following Buterin, we assume that ``we generally can rely on assumptions that at least a few percent of proposers are altruistic''~\cite{buterin2021design}.
This assumption drives recent fairness mechanisms, most notably \textsc{Inclusion Lists} (ILs)~\cite{EIP7547,thiery2024eip7805} and \textsc{Multiple Concurrent Proposer} (MCP) designs~\cite{Garimidi2025concurrent}, which aim to mitigate temporary censorship~\cite{censorship}.
These mechanisms replace single-proposer control with committee-based control over block contents.
For these mechanisms to work as intended, each committee must contain at least one proposer who enforces the inclusion of a transaction targeted for censorship, even when financial incentives, such as bribes, favor exclusion~\cite{berger2025economiccensorshipgamesfraud}.
Their effectiveness therefore depends on two parameters: the committee size, a design choice with limited scalability, and the share of altruistic proposers, which -- to the best of our knowledge -- has not been empirically quantified.
This gap motivates our research question: \emph{What share of Ethereum proposers can be identified as exhibiting altruistic behavior?}

\vspace{2.2mm}
Prior work~\cite{heimbach_ethereums_2023} has shown that approximately \SI{90}{\percent} of blocks are constructed by centralized block construction services~\cite{MEV_definition_Sedlmeir}, thereby causing the respective proposers to relinquish control over block contents (cf. \Cref{sec:on_alt_mev}).
During our observation period (Jan.--Dec. 2025), \SI{91.3}{\percent} of blocks were published via such services.
If one assumes that \SI{8.7}{\percent} of proposers are altruistic, then the probability that a 16-member committee, as proposed in Ethereum's upcoming Inclusion List 
design~\cite{thiery2024eip7805},\footnote{Fork-choice enforced Inclusion Lists (FOCIL)~\cite{thiery2024eip7805} is scheduled for deployment in Ethereum's Hegota Upgrade~\cite{ethereum_hegota_2026}, expected by the end of 2026.} contains at least one altruist is $1 - 0.913^{16} \approx \SI{76.7}{\percent}$.
This suggests a substantial improvement in expected censorship resistance.
Our measurements, however, suggest that at most \SI{1.55}{\percent} of proposers can be classified as altruistic based on observable behavior.
Under this share, the same calculation yields $1 - 0.9845^{16} \approx \SI{22.1}{\percent}$, which, from our perspective, warrants further discussion.

\vspace{1mm}
\paragraph*{Contributions}
Through an empirical analysis, we estimate the share of altruistic proposers on Ethereum and assess the implications for fairness mechanisms that rely on altruistic participation.
Starting from a set of roughly one million proposers, we progressively exclude those for whom we observe unambiguous evidence of non-altruistic behavior.
The complementary set yields an upper bound, which we describe as the set of ``potentially altruistic'' proposers.
Using stricter exclusion criteria, we further identify proposers for whom we do not observe any evidence of deviation from the protocol and interpret this set as a lower bound on the share of genuinely altruistic proposers.
Our findings place the share of altruistic proposers in the interval $[\SI{0.41}{\percent};\SI{1.55}{\percent}]$.

\vspace{2mm}
The paper is organized as follows: In \Cref{sec:altruism}, we outline our rationale for the classification of altruism and illustrate non-altruistic behaviors.
\Cref{sec:methodology} outlines the data sources and the classification methodology, followed by quantitative findings in \Cref{sec:results}.
\Cref{sec:discussion} discusses these findings with respect to their relevance for fairness mechanisms as well as the employed measurement methodology.
\Cref{sec:conclusion} concludes the paper.
Related work is referenced inline.

\section{Defining Non-Altruistic Behavior}
\label{sec:altruism}
The term ``altruism'' is conceptually ambiguous and difficult to ground empirically, since underlying motivations are not observable on-chain.
Our objective is not to assess proposer ethics, but to gauge whether a proposer would deviate from a fairness mechanism once higher revenues become attainable.
We therefore adopt a strictly behavior-based definition: altruistic behavior is understood as proposer behavior that adheres to Ethereum's intents without deviation.
Deviations manifest along two observable dimensions: MEV extraction or its delegation to external builders, and shared account governance -- both of which frequently co-occur in practice.
The following sections refine our behavioral definition along each dimension.

\subsection{Extracting Maximal Extractable Value}
\label{sec:on_alt_mev}
MEV refers to the additional profit a proposer can capture by strategically ordering, including, or excluding transactions within or across blocks~\cite{MEV_definition_Sedlmeir}.
By observing the mempool, a local storage for pending transactions, proposers or delegated entities can exploit publicly broadcast transactions through techniques such as front-running or sandwich attacks~\cite{MEV_Attacks,daian_flash_2020}.
In addition, MEV-related transaction censorship can also occur, for example by accepting bribes~\cite{berger2025economiccensorshipgamesfraud} to censor transactions or by preventing the liquidation of own assets~\cite{NADLER2026102908,wu2024strategicbiddingwarsonchain}.
These strategies distort execution outcomes and impose additional costs on users, as investigated in prior work~\cite{daian_flash_2020}.
We define MEV extraction as non-altruistic behavior when profit-maximizing strategies exploit transaction contents instead of relying on protocol-defined inclusion and prioritization mechanisms, such as transaction priority fees~\cite{EIP1559}.

In principle, a proposer could extract MEV directly.
However, in practice, outsourcing this task to specialized entities, called builders, is often the financially preferable choice.
Flashbots' MEV-Boost~\cite{flashbots_mev_boost} exemplifies such delegation by acting as trusted middleware, with its relays mediating block contents between proposers and builders, thereby implementing Proposer-Builder Separation (PBS).\footnote{As of 2025, EIP-7732~\cite{EIP7732} (enshrined PBS at the consensus layer, ePBS) was not deployed on Ethereum. We discuss its impact in \Cref{sec:discussion}.}
In this setup, the builder constructs the block with the objective of maximizing profit, while the proposer merely signs the block selected by an MEV-Boost relay without any visibility into its contents. 
Typically, the relay selects the block with the highest bid.
Upon receiving the proposer's signature, the relay propagates the block.
As a result, the proposer cannot inspect or modify the block's transactions prior to committing, effectively ceding control over transaction inclusion, exclusion, and ordering to the builder.
This setup centralizes decision-making in the hands of successful builders, who remain competitive by using profit-maximizing strategies that could disadvantage users.
Empirically, MEV-Boost dominates Ethereum today.
Between Q3 2023 and Q4 2025, \SI{88.1}{\percent}--\SI{94.3}{\percent} of daily blocks were mediated via MEV-Boost relays~\cite{heimbach_ethereums_2023}.
Furthermore, the two largest builders construct over \SI{80}{\percent} of all blocks~\cite{heimbach_ethereums_2023}, illustrating the high degree of centralization in decision-making.

\vspace{1mm}
Consequently, while PBS can be regarded as part of Ethereum's intended design~\cite{ethereum_pbs}, its realization through MEV-Boost deviates from this design~\cite{heimbach_ethereums_2023}.
In particular, proposers retain no meaningful control over block contents -- for instance, the ability to prevent transaction censorship.
The fairness mechanisms discussed above rely on proposers being willing to retain such control, for example by enforcing the inclusion of transactions despite financial incentives to censor them.
Such incentives may arise as bribes paid to exclude transactions~\cite{berger2025economiccensorshipgamesfraud}, or more generally through implicit agreements that grant builders the freedom to omit certain transactions.
We therefore argue that proposers who either pursue utility-maximizing and user-adverse strategies or fully outsource block construction for profit are likely to prioritize profit over protocol intent whenever deviation is financially attractive, and should thus not be classified as altruistic, but rather as rational.
Accordingly, such proposers should not be relied upon in fairness mechanisms under a 1-of-\textit{n} honest assumption.

\subsection{Sharing Account Governance}
\label{sec:on_alt_gov}
We refer to the entity that provides the financial resources for a validator as the stakeholder.\footnote{\label{footref:staking} A validator account must deposit at least 32 Ether, Ethereum's native cryptocurrency, which corresponded to over \$\num{115000} as of January 2025.}
Operational control over a validator follows exclusively from possession of the validator's private key, which, under Ethereum's ideal, is held by individual stakeholders.
Running a validator, however, requires highly available infrastructure for frequent duties as well as ongoing maintenance to remain compatible with protocol upgrades.
Due to these requirements, stakeholders typically share infrastructure~\cite{heimbach_ethereums_2023} that holds the private keys of multiple validators and exercises operational control over them.
Granting an infrastructure provider access to a validator's private key requires the stakeholder to trust the provider, operate the infrastructure themselves, or rely on an externally managed custodial arrangement.
Given the substantial capital locked in validator accounts,\footref{footref:staking} blind trust is implausible.
This leaves two prevalent configurations.
First, stakeholders can operate their own infrastructure and manage their validators themselves, implying that a single infrastructure setup holds the private keys for all associated validators.
Second, stakeholders may outsource validator operation to external infrastructures with external custodial arrangements, which retain custody of the private keys for validators of multiple stakeholders.
Such entities are commonly referred to as Staking-as-a-Service (STaaS) providers~\cite{ethereum_saas}, including staking pools.
For our purposes, the relevant unit of analysis is thus not always the individual validator.
What matters is the entity that holds operational control and therefore determines block-construction decisions.
In both self-managed and externally managed infrastructure setups, we refer to this entity as the managing party.
The managing party, together with the validators it controls, forms a governance structure~\cite{ryan2022lsd}.
Accordingly, we define shared governance as any arrangement in which one managing party controls the keys of at least two validators, regardless of who the stakeholders are.
This follows directly: once operational control is centralized at the level of the managing party, behavioral assumptions should also be made at that level.
If a managing party demonstrates non-altruistic behavior through one proposer, it is reasonable to treat the other proposers under the same control as exposed to the same incentives, policies, and operational decisions.
Accordingly, we do not rely solely on block-level evidence for non-altruistic behavior, as is typically done in related work.

The following two examples illustrate why isolated block-level observations are insufficient to characterize proposer behavior.
First, proposers can set instructions for MEV-Boost relays, specifying that blocks from a builder are only accepted if the builder's offer exceeds a predefined threshold compared to the rewards for constructing the blocks themselves.\footnote{In Prysm~\cite{prysm_builder_config}, a widely used consensus client developed by Offchain Labs, this setting is called \colorbox{gray!20}{\texttt{local-block-value-boost}}. Offchain Labs states that this mechanism mitigates censorship concerns.\label{footnote:partAlt}}
When a block offers only limited MEV extraction opportunities, this instruction may be triggered, making the proposer appear indistinguishable from an altruistic one.
Yet the relevant decision rule remains non-altruistic: the proposer has opted into profit-maximizing outsourcing and would have accepted the builder's block if doing so had been sufficiently profitable.
Given a sufficiently large bribe, we would therefore expect this proposer to behave identically to the other proposers within the same governance structure, including a willingness to deviate from a fairness mechanism whenever doing so is financially beneficial.
The apparently altruistic outcome is contingent on circumstances, not on an altruistic policy.
\enlargethispage{2\baselineskip}
Second, decentralized STaaS networks such as Lido~\cite{Lido2025} operate with a rotating set of managing parties rather than a single fixed operator.
As a result, operational control over a validator is time-dependent and may alternate among multiple managing parties.
Since stakeholders have no influence over which party controls their validator at any given time, participation in such a network implicitly entails acceptance of rational, non-altruistic behavior whenever at least one managing party acts non-altruistically.
Observing altruistic behavior for a proposer at a specific point in time is therefore insufficient to classify that proposer as altruistic.
At other times, and without the possibility of stakeholder intervention, the same proposer may be controlled by a non-altruistic managing party.
We therefore argue that, in shared governance setups with a predominantly rational managing party, any proposer under that governance structure may deviate from a protocol at an unpredictable point in time whenever such a deviation is financially beneficial.

\raggedbottom
\section{Methodology}
\label{sec:methodology}
To quantify the prevalence of altruistic proposer behavior, we conducted an empirical study over the full year 2025, evaluating each quarter individually to confirm the stability of our findings.
Our methodology employs a multi-stage classification pipeline.
First, we exclude proposers for whom we observe clear evidence of non-altruistic behavior, thereby yielding an upper bound on the altruistic set.
Second, we identify proposers for whom altruistic behavior is unambiguously observable during the observation period, yielding a lower bound.
We draw on four data sources.
From Ethereum's execution layer, accessed via a self-operated node, we extract the transactions included in each block, recording the priority fees paid by users and the \emph{coinbase address} that identifies the account receiving these fees.
From the Beacon Chain (consensus layer), accessed through the Beaconcha.in~\cite{beaconchain_api} API, we identify the protocol-assigned proposer~\cite{ETHspec2} via its proposer index.
MEV-Boost usage is determined from the public endpoints of well-known relays, which publish information about their mediated blocks~\cite{flashbots_relay_specs}.
We further incorporate network-layer data from 
\cite{mempool_guru} to identify non-broadcast transactions.
\begin{figure*}[!ht]
    \centering
    \includegraphics[trim={0cm 21.7cm 0cm 0.5cm}, clip, width=0.72\textwidth,page=1]{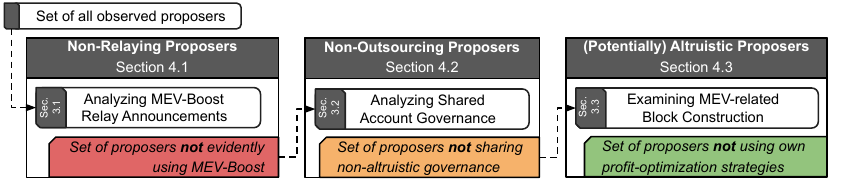}
    \caption{Methodology for identifying altruistic proposers: starting from all observed proposers, excluding evident MEV-Boost participants (cf. \Cref{sec:on_alt_mev}), checking shared account governance (cf. \Cref{sec:on_alt_gov}), and detecting block-level MEV-related block construction (cf. \Cref{sec:on_alt_mev}). \Crefrange{sec:3.1}{sec:3.3} refer to methodology, \Crefrange{sec:4.1}{sec:4.3} to results.}
    \label{fig:methodology-overview}
    \vspace{-0.3cm}
\end{figure*}

\vspace{1mm}
Our method (visualized in \Cref{fig:methodology-overview}) consists of three steps that employ six classifiers to identify non-altruistic proposers and exclude them from further consideration.
First, we exclude proposers identified as relying on MEV-Boost relays, classified as \textit{relaying proposers}, since they blindly sign externally constructed blocks.
The set of proposers not evidently using MEV-Boost remains (marked in red).
Second, we assess whether the remaining \textit{non-relaying} proposers participate in shared governance structures.
We exclude those proposers who, despite not being announced by MEV-Boost relays, share a governance structure with relaying proposers.
These proposers are classified as \textit{outsourcing proposers}.
The set of proposers not sharing non-altruistic governance remains (marked in orange).
Third, within the set of non-outsourcing proposers, we analyze proposed blocks to determine whether MEV-related strategies were applied, even in the absence of detectable interaction with external builders.
Details are provided below.
Proposers are classified as altruistic if they neither use MEV-Boost relays nor participate in shared governance with already excluded proposers, and do not execute MEV extraction.

\raggedbottom
\subsection{Analyzing Relay Announcements}
\label{sec:3.1}
Classifier \Circled{\RNum{1}} distinguishes between proposers whose blocks were relayed via MEV-Boost~\cite{flashbots_mev_boost} and those that were not.
We exclude any proposer whose blocks were mediated by an MEV-Boost relay at least once.
This constitutes the most evident form of outsourcing and leads to exclusion from the altruistic set, as it leaves the proposer with no means to counteract unfair block construction.

\vspace{1mm}
\textsc{Method \textup{\Circled{\RNum{1}}}:}
To identify \textit{relaying proposers}, we obtain mediated blocks via the public JSON-RPC endpoint (\colorbox{gray!20}{\texttt{proposer\_}}\allowbreak\colorbox{gray!20}{\texttt{payload\_}}\allowbreak\colorbox{gray!20}{\texttt{delivered}}) of MEV-Boost relays.
Each mediated block is linked to its proposer using Beacon Chain data.
This procedure allows us to compute both the fraction of mediated blocks among all observed blocks and the fraction of proposers that used MEV-Boost at least once.

\subsection{Analyzing Shared Account Governance}
\label{sec:3.2}
After excluding proposers who clearly rely on external builders, we next exclude proposers for whom we find evidence of participation in a shared governance structure.
Building on our rationale in \Cref{sec:on_alt_gov}, Classifier \Circled{\RNum{2}} groups proposers by governance structure and treats them as belonging to the same behavioral unit.
If a cluster contains at least one non-altruistic proposer, all proposers in this cluster are excluded.

\vspace{1mm}
\textsc{Method \textup{\Circled{\RNum{2}}}:} We cluster proposers by the \emph{coinbase address} recorded in their proposed blocks; a similar approach, with a different research focus, was taken by Grandjean et al.~\cite{grandjean_participation}.
This yields a mapping from \emph{coinbase addresses} to the set of proposers publishing blocks with the same transaction tip recipient.
We consider the probability that a proposer would send its rewards to a \emph{coinbase address} without direct access or an explicit agreement with its managing party as negligible, justifying the use of the address as a cluster identifier.

\vspace{1mm}
The clustering proceeds as follows.
Consider two proposers $P_A$ and $P_B$ that publish blocks with the same \emph{coinbase address}.
If $P_A$'s block is announced via an MEV-Boost relay while $P_B$'s is not, we classify $P_B$ as non-altruistic due to its shared governance with $P_A$, who is already excluded as a relaying proposer.
The absence of a relay announcement of $P_B$'s block is attributed to other factors, such as incorrect relay reporting or strategic profitability choices.\footref{footnote:partAlt}
If no other proposer shares the \emph{coinbase address} of a proposer $P_C$, the induced cluster has size one and contains only $P_C$.

\vspace{2mm}
\noindent As a further refinement step, using Classifier \textup{\Circled{\RNum{3}}}, we search for payments from well-known builders~\cite{wahrstatter_time_2023} to the \textit{coinbase addresses} of the remaining clusters.
These payments correspond to funds that builders transfer to proposers in exchange for the right to determine their block contents.
According to our rationale in \Cref{sec:on_alt_mev,sec:on_alt_gov}, proposers in these clusters are excluded from the altruistic set.
This behavior is particularly prevalent among MEV-extracting STaaS providers, further supporting the classification of their proposers as non-altruistic.

\enlargethispage{1\baselineskip}
\vspace{1mm}
\textsc{Method \textup{\Circled{\RNum{3}}}:}
To identify these payments, we query the MEV-Boost relay endpoints\footref{considered_relays} for blocks published by proposers who have not been excluded so far.
The relay response includes a field named \colorbox{gray!20}{\texttt{proposer\_}}\allowbreak\colorbox{gray!20}{\texttt{fee\_}}\allowbreak\colorbox{gray!20}{\texttt{recipient}},\footnote{This field is provided by the MEV-Boost protocol and does not correspond to the block's specified \colorbox{gray!20}{\texttt{fee\_}}\allowbreak\colorbox{gray!20}{\texttt{recipient}} (\textit{coinbase address}).} which specifies the address receiving the builder's winning bid.
If an MEV-Boost relay returns a recipient matching an unexcluded \textit{coinbase address} of a cluster, we exclude all proposers within that cluster from the altruistic set.

\newpage
\subsection{Examining Block Construction}
\label{sec:3.3}
In \Cref{sec:3.1,sec:3.2}, we examined profit-maximizing behavior delegated to external builders.
The following classifiers address MEV extraction performed directly by proposers, analyzing two degrees of freedom identified in prior work~\cite{MEV_definition_Sedlmeir}: the inclusion of non-publicly propagated transactions (Exclusive Order Flow, XOF) and the ordering of transactions within a block.
The lenient Classifier \textup{\Circled{\RNum{4}}} excludes proposers with conclusive evidence of non-altruistic behavior; the remaining set constitutes our upper bound on altruistic proposers.
Classifiers \textup{\Circled{\RNum{5}}} and \textup{\Circled{\RNum{6}}} together impose the strictest interpretation of protocol-compliant behavior, with Classifier \textup{\Circled{\RNum{6}}} yielding the lower bound.
We therefore expect the genuine share of altruistic proposers to fall between the upper bound from Classifier \textup{\Circled{\RNum{4}}} and the lower bound from Classifier \textup{\Circled{\RNum{6}}}.\\

\vspace{-3mm}
Classifiers \textup{\Circled{\RNum{4}}} and \textup{\Circled{\RNum{5}}} examine XOF transactions, whose inclusion constitutes evidence of direct MEV extraction by the proposer (cf. \Cref{sec:on_alt_mev}).
Private order flow has been shown to account for more than half of extractable MEV value~\cite{wang2025privateorderflows} and relies on private communication channels outside of Ethereum's intended transaction propagation model.
Classifier \textup{\Circled{\RNum{4}}} excludes proposers whose blocks contain more than one transaction not visible in our dataset; manual inspection suggests that a single unobserved transaction per block is more likely a data artifact in the mempool dataset than genuine MEV-related behavior.
The strict Classifier \textup{\Circled{\RNum{5}}} excludes any proposer with at least one XOF transaction, following the rationale of \Cref{sec:on_alt_mev} and yielding a tighter candidate set.

\textsc{Method \textup{\Circled{\RNum{4}}} and \textup{\Circled{\RNum{5}}}:}
To detect XOF, we examine the transactions included in blocks proposed by the remaining proposers and check whether they appear in the dataset provided by \cite{mempool_guru}.
This dataset contains all propagated transactions observed by at least one of four monitoring nodes.
Transactions can therefore be divided into those present in the dataset and those absent, the latter of which we attribute to XOF.
While the dataset covers four monitoring nodes, cross-checking against an additional monitor (Etherscan~\cite{etherscan}) reveals minor discrepancies.
To avoid underestimating the share of altruistic proposers, we exclude only the identified proposer rather than the entire cluster.

\vspace{1mm}
Classifier \textup{\Circled{\RNum{6}}} examines whether a block's transaction ordering is verifiably content-independent, i.e., explainable solely by publicly observable attributes.
A proposer who neither includes XOF transactions nor deliberately reorders transactions cannot extract MEV in the typical sense; a verifiably content-independent ordering provides evidence that no such reordering occurred.
Note, however, that failing this criterion does not imply non-altruistic behavior: some orderings are content-independent but not externally verifiable, e.g., when ordered by internal arrival time or to respect account-nonce dependencies.
Classifiers \textup{\Circled{\RNum{5}}} and \textup{\Circled{\RNum{6}}} together form the boundary between the \emph{potentially altruistic} and the \emph{altruistic} set, with Classifier \textup{\Circled{\RNum{6}}} serving as the final refinement.
Since exclusion under Classifier \textup{\Circled{\RNum{6}}} could reflect a lack of verifiability rather than observed non-altruistic behavior, we exclude only individual proposers rather than entire clusters.

\vspace{1mm}
\textsc{Method \textup{\Circled{\RNum{6}}}:}
We investigate whether transactions within blocks are ordered according to publicly observable, protocol-defined attributes.
Empirically, higher-fee transactions tend to appear earlier, resulting in an approximately descending sequence of \colorbox{gray!20}{\texttt{maxPriorityFeePerGas}}, consistent with Ethereum's intended fee model and verifiable by all participants~\cite{EIP1559}.
To quantify ordering beyond a binary distinction, we compute the Spearman rank correlation coefficient $\sigma_{rs}$\cite{spearman} between transaction priority fees and their positions within the block.
We treat $\sigma_{rs} = 1$ as our strict ordering criterion, as it unambiguously identifies blocks ordered solely by priority fee.

\section{Results}
\label{sec:results}
\begin{figure}[!ht]
    \centering
    \includegraphics[trim={0cm 16cm 2cm 0cm}, clip, width=\linewidth,page=2]{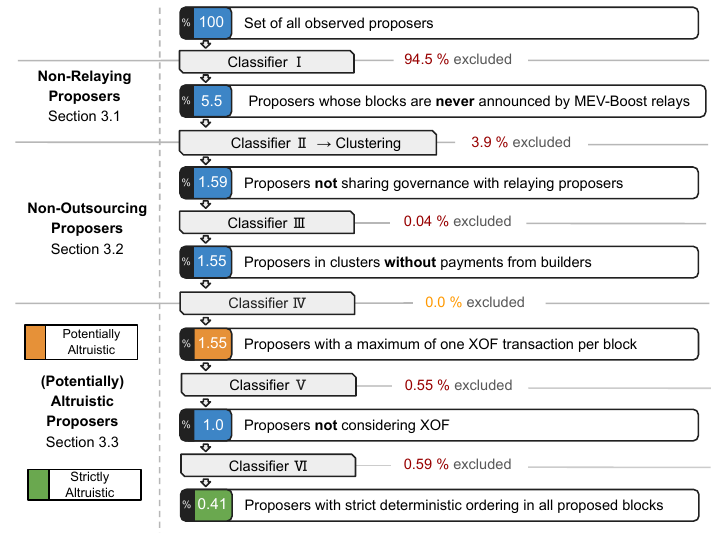}
    \caption{Analysis results -- percentages denote shares of the observed proposer set (1\,108\,992 proposers in total).}
    \Description{Overview diagram of classifier results and the remaining proposer shares after each exclusion step.}
    \label{fig:results-overview}
\end{figure}

In this section, we present the results obtained using the classifiers introduced in \Cref{sec:methodology}.
We report results for the full measurement period (Q1--Q4 2025), summarized in \Cref{fig:results-overview}.
Additionally, we evaluate each quarter individually to assess the temporal stability of our findings; these results are reported in \Cref{tab:proposers} and discussed in \Cref{discussion:observation_window}.
By construction, the strictly altruistic set is a proper subset of the potentially altruistic set.

\begin{table*}[ht]
\centering
\begin{tabular}{lrrrrrrrrrrrr}
\toprule
& \multicolumn{3}{c}{\textbf{Q1 2025}} & \multicolumn{3}{c}{\textbf{Q2 2025}} & \multicolumn{3}{c}{\textbf{Q3 2025}} & \multicolumn{3}{c}{\textbf{Q4 2025}} \\
\cmidrule(lr){2-4} \cmidrule(lr){5-7} \cmidrule(lr){8-10} \cmidrule(lr){11-13}
\textbf{Stage} & \textbf{Excl.} & \textbf{Rem.} & \textbf{Rem.\%} & \textbf{Excl.} & \textbf{Rem.} & \textbf{Rem.\%} & \textbf{Excl.} & \textbf{Rem.} & \textbf{Rem.\%} & \textbf{Excl.} & \textbf{Rem.} & \textbf{Rem.\%} \\
\midrule
All proposers & -- & \num{487435} & 100.00 & -- & \num{489762} & 100.00 & -- & \num{487221} & 100.00 & -- & \num{455875} & 100.00 \\[2mm]
Classifier \textup{\Circled{\RNum{1}}} & \num{446619} & \num{40816} & 8.37 & \num{454035} & \num{35727} & 7.29 & \num{452927} & \num{34294} & 7.04 & \num{418003} & \num{37872} & 8.31 \\[2mm]
Classifier \textup{\Circled{\RNum{2}}} & \num{29392} & \num{11424} & 2.34 & \num{26390} & \num{9337} & 1.91 & \num{25421} & \num{8873} & 1.82 & \num{29572} & \num{8300} & 1.82 \\[2mm]
Classifier \textup{\Circled{\RNum{3}}} & \num{1093} & \num{10331} & 2.12 & \num{1113} & \num{8224} & 1.68 & \num{870} & \num{8003} & 1.64 & \num{813} & \num{7487} & 1.64 \\[2mm]
Classifier \textup{\Circled{\RNum{4}}} & \num{0} & \textcolor{orange}{\num{10331}} & \textcolor{orange}{2.12} & \num{0} & \textcolor{orange}{\num{8224}} & \textcolor{orange}{1.68} & \num{0} & \textcolor{orange}{\num{8003}} & \textcolor{orange}{1.64} & \num{0} & \textcolor{orange}{\num{7487}} & \textcolor{orange}{1.64} \\[1mm]
Classifier \textup{\Circled{\RNum{5}}} & \num{2354} & \num{7977} & 1.64 & \num{1759} & \num{6465} & 1.32 & \num{1713} & \num{6290} & 1.29 & \num{1457} & \num{6030} & 1.32 \\[2mm]
Classifier \textup{\Circled{\RNum{6}}} & \num{4106} & \textcolor{green!60!black}{\num{3871}} & \textcolor{green!60!black}{0.79} & \num{3894} & \textcolor{green!60!black}{\num{2571}} & \textcolor{green!60!black}{0.52} & \num{3219} & \textcolor{green!60!black}{\num{3071}} & \textcolor{green!60!black}{0.63} & \num{2999} & \textcolor{green!60!black}{\num{3031}} & \textcolor{green!60!black}{0.66} \\
\bottomrule
\end{tabular}
\caption{Analysis results for slot ranges Q1 [10\,738\,799; 11\,386\,798], Q2 [11\,386\,799; 12\,041\,998], Q3 [12\,041\,999; 12\,704\,398], and Q4 2025 [12\,704\,399; 13\,366\,798]. \textbf{Excl.} and \textbf{Rem.} denote the number of proposers excluded by, and remaining after, the respective classifier. Orange-highlighted values denote the potentially altruistic set, while green-highlighted values denote the altruistic set. Percentages are relative to all proposers (\textbf{100\,\%}).
Note that within the full measurement period, the strictly altruistic share is lower than in any individual quarter, since there is no possibility that an excluded proposer can be reclassified as altruistic. We discuss the choice of the observation windows in~\Cref{sec:discussion}.
}
\label{tab:proposers}
\vspace{-0.3cm}
\end{table*}

\vspace{-3mm}
\subsection{Quantifying Non-Relaying Proposers}
\label{sec:4.1}
During the observation period, we analyzed \num{2610762} consecutive blocks in the slot range [\num{10738799}; \num{13366798}], excluding skipped slots.
These blocks were proposed by \num{1108992} distinct proposers.
Of all blocks, \num{2383636} blocks (\SI{91.3}{\percent}) were announced by at least one relay,\footnote{We considered the following well-known relays: Aestus, Agnostic, bloXroute (Regulated and Max-Profit), Flashbots, Ultrasound, and Titan.\label{considered_relays}} in line with previous studies~\cite{heimbach_ethereums_2023,wahrstatter_time_2023,yang_decentralization_2025}.
We distinguish three groups based on their relay usage:
\begin{enumerate}[label=(\roman*)]
\item \num{964951} (\SI{87}{\percent}) distinct proposers whose blocks were exclusively announced via relays,
\item \num{83150} (\SI{7.5}{\percent}) distinct proposers with a proper subset of blocks announced via relays, and
\item \num{60891} (\SI{5.5}{\percent}) distinct proposers whose blocks (median: two blocks) were never announced via relays.
\end{enumerate}
With regard to Classifier \textup{\Circled{\RNum{1}}}, we restrict our analysis to the \num{60891} non-relaying proposers whose blocks were never announced by known MEV-Boost relays.
This excludes \SI{94.5}{\percent} of proposers that used block-building services at least once.

\subsection{Quantifying Non-Outsourcing Proposers}
\label{sec:quantifying_nop}
\begin{figure}[t]
    \centering
    \includegraphics[trim={1.5cm 0cm 1cm 1.3cm}, clip, width=\linewidth]{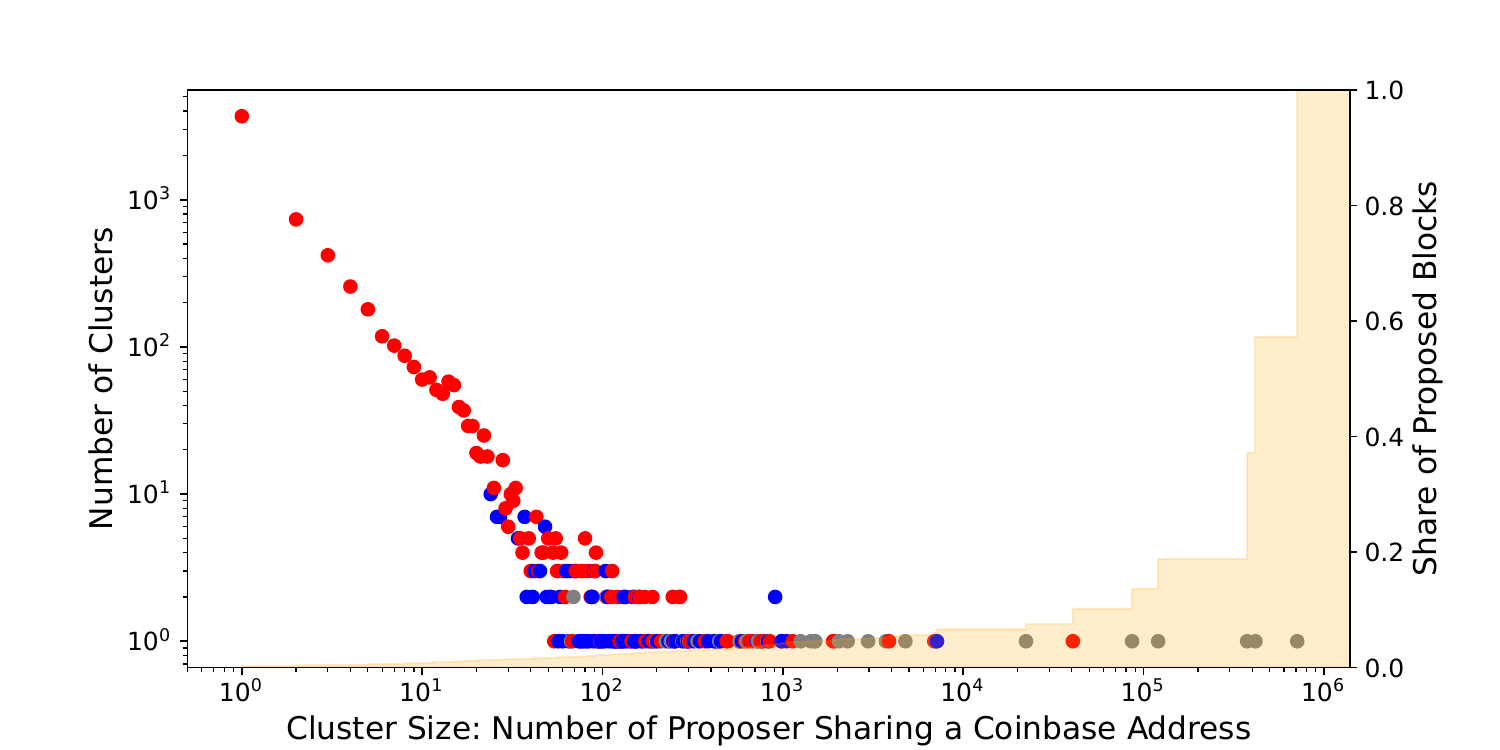}
        \caption{Histogram of number of proposers sharing a coinbase address (clusters).
        Grey dots denote clusters in which all proposers' blocks were announced by an MEV-Boost relay, excluded by Classifier \textup{\Circled{\RNum{1}}}. Blue dots indicate clusters in which no proposer published a block announced by any MEV-Boost relay. Red dots indicate clusters (not necessarily all of the same size) that include a proper subset of proposers using MEV-Boost. These proper subsets are excluded under Classifier \textup{\Circled{\RNum{2}}}. The cumulative distribution function shows the relative share of blocks produced by clusters, highlighting that three clusters (attributable to the two largest builders~\cite{heimbach_ethereums_2023}) account for more than \textbf{80\,\%} of all blocks.}
    \label{fig:coinbase-distribution}
    \vspace{-0.1cm}
\end{figure}

To quantify proposers whose blocks were not announced via MEV-Boost relays, but for whom links to external builder services can still be observed, we apply our clustering methodology to the full proposer set.
\Cref{fig:coinbase-distribution} summarizes the resulting cluster sizes and frequencies based on \textit{coinbase addresses}.
In what follows, we remove clusters rather than individual proposers.
Applying Classifier \textup{\Circled{\RNum{2}}} (non-altruistic governance structures) reduces the set to \SI{1.59}{\percent}.
Applying Classifier \textup{\Circled{\RNum{3}}} (block builder payments) further reduces the upper bound on altruistic proposers to \SI{1.55}{\percent}.
These \SI{1.55}{\percent} show no evidence of outsourced block construction.
To assess the impact of potential misclassifications, we note that no cluster of size $\geq 8$ is excluded based on only a minority of its proposers.

\vspace{1mm}
In summary, while \SI{91.3}{\percent} of blocks are mediated by MEV-Boost relays, we observe evidence of interactions with external builders for \SI{98.45}{\percent} of proposers, indicating that \SI{98.45}{\percent} of proposers can be classified as demonstrably non-altruistic.
With respect to fairness mechanisms, such proposers cannot be assumed to consistently act in a manner conducive to these mechanisms.
They are likely to deviate whenever deviation is the rational choice -- for instance, by delegating their decision-making power to a rational entity.

\subsection{Quantifying Potentially Altruistic Proposers}
\label{sec:4.3}
Having excluded proposers involved in profit-maximizing activities through external services, we turn to proposers who may extract MEV directly.
Classifier \textup{\Circled{\RNum{4}}} yields no additional exclusions; the potentially altruistic set thus remains at \SI{1.55}{\percent} of all observed proposers.
Applying the stricter Classifier \textup{\Circled{\RNum{5}}}, which excludes any proposer with at least one XOF transaction, reduces the set to \SI{1.0}{\percent}.
Classifier \textup{\Circled{\RNum{6}}} identifies the altruistic set: in total, \SI{0.41}{\percent} of proposers satisfy all criteria, including our strict ordering criterion ($\sigma_{rs}=1$).
For the proposers excluded under this classifier, the ordering of their blocks is not strictly consistent, yet it remains strongly aligned with priority fees: \SI{99.3}{\percent} exhibit a Spearman coefficient $\sigma_{rs} > 0.7$, while \SI{35.7}{\percent} reach $\sigma_{rs} > 0.99$.
Several proposers fail the strict-ordering criterion despite consistently high Spearman correlations across all blocks, which may be attributed to user-specified transaction ordering or other block optimizations unrelated to MEV activity, indicating that Classifier \textup{\Circled{\RNum{6}}} underestimates the genuine share of altruistic proposers.

\subsection{Characterizing Altruistic Proposers}
After applying Classifiers \textup{\Circled{\RNum{1}}}--\textup{\Circled{\RNum{4}}}, \num{17144} proposers (\SI{1.55}{\percent} of the total set) show no conclusive evidence of non-altruism and are therefore classified as potentially altruistic.
\num{4598} proposers (\SI{0.41}{\percent}), classified as strictly altruistic, exhibit no indication of deviating from Ethereum's intended behavior under any of our criteria.
Cluster analysis indicates that potentially altruistic proposers are typically organized into small governance structures: the median cluster contains a single proposer, while some reach over \num{900} members.

\begin{figure}[!h]
    \centering
    \includegraphics[trim={0.2cm 1.2cm 1cm 2cm}, clip, width=0.95\linewidth]{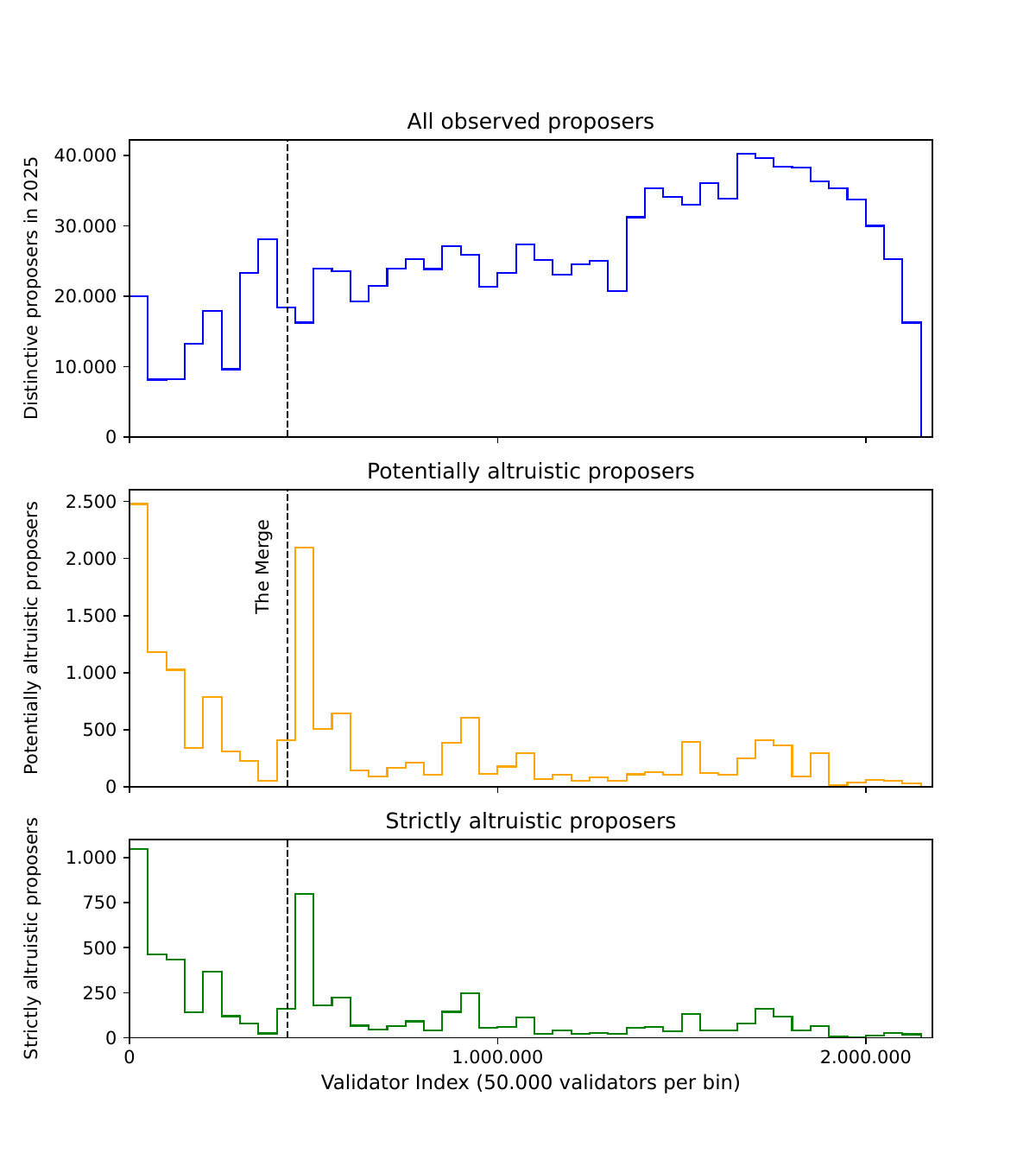}
    \caption{Histogram of proposers classified as altruistic (green) and potentially altruistic (orange), together with the number of proposers that proposed at least one block during our full measurement period (blue). Each bin covers 50 000 validators; the dashed line separates proposers registered before and after Ethereum's ``The Merge''~\cite{ethereum2024merge}.}
    \label{fig:potentially_altruistic_timeline}
    \vspace{-0.5cm}
\end{figure}

While we do not attempt to de-anonymize these larger clusters for ethical reasons, we provide additional context for the (potentially) altruistic set through a temporal observation.
A substantial share of potentially altruistic proposers (\SI{44}{\percent}) registered before Ethereum's transition to Proof-of-Stake (``The Merge'')~\cite{ethereum2024merge}; among strictly altruistic proposers, the share is \SI{45.9}{\percent}:
these proposers were active during the network's test phases, before transaction fees and MEV could be extracted.
\Cref{fig:potentially_altruistic_timeline} shows the entire proposer set alongside the (potentially) altruistic subset.
While newer proposers announce more blocks overall -- as many older validators are no longer active -- potentially altruistic behavior occurs at disproportionately higher rates among older cohorts: \SI{15}{\percent} of proposers in the validator-index range $[\num{0};\num{100000})$ are classified as potentially altruistic, compared to \SI{0.59}{\percent} in the most recent range $[\num{1000000};~\num{2000000})$.
These findings suggest that early-phase proposers, who participated before protocol-deviating economic incentives existed, are overrepresented in the (potentially) altruistic set -- consistent with a participation motive that does not primarily aim at maximizing revenue.

\section{Discussion}
\label{sec:discussion}
\paragraph*{On Fairness Mechanisms}
Both ILs~\cite{thiery2024eip7805} and MCP designs~\cite{Garimidi2025concurrent} place the decision over mandatory transaction inclusion with a committee rather than a single proposer.
Since committee members typically pursue heterogeneous objectives and operate under different governance constraints~\cite{censorship}, censoring a transaction requires all members to agree on the exclusion of that transaction.
Bribes~\cite{berger2025economiccensorshipgamesfraud} can align these otherwise heterogeneous objectives; robustness then rests on at least one member refusing the bribe -- that is, on altruistic behavior.
Note that the ``honest'' member in the 1-of-$n$ assumption coincides with our notion of an altruistic proposer: we model rational proposers as fully rational, accepting even arbitrarily small bribes whenever deviation is profitable.
The practical realization of bribing committee members into protocol deviations~\cite{bribers} with small incentives warrants further investigation.
Under our notion of altruism, altruistic proposers do exist but constitute at most a \SI{1.55}{\percent} share, almost an order of magnitude below the \(\approx\)\,\SI{9}{\percent} assumption.
For altruism-dependent censorship-mitigation mechanisms, this implies the need for substantially larger committees.
Ensuring that a committee contains at least one altruistic member with a probability of at least \SI{80}{\percent} requires more than \num{100} members -- for instance, \num{128} members yield \(1 - 0.9845^{128} \approx\)~\SI{86.4}{\percent} -- compared to the \num{16}-member committees originally proposed~\cite{thiery2024eip7805}.
At this scale, individual members exert correspondingly less influence over the resulting block, raising concerns about the practical viability of mechanisms that depend solely on altruistic participation.
Several recent designs therefore replace the reliance on altruism by aligning fair behavior with rational incentives.
Fox et al.~\cite{fox2023censorship} reward proposers for transaction inclusion, paying more for transactions endorsed by fewer committee members.
Wadhwa et al.\ introduce \textsc{AUCIL}~\cite{aucil}, an IL design in which inclusion is the rational default for committee members, so that excluding a transaction requires a bribe several orders of magnitude higher than bribes sufficient today.
Spiesberger et al.~\cite{spiesberger_policybased_2025} follow an accountability-based approach: the builder must either include every transaction whose priority fee warrants inclusion or publicly disclose each omission, thereby becoming accountable for every censored transaction.
A committee verifies the completeness of these disclosures rather than prescribing block contents -- a passive role that permits substantially larger committees.
Undisclosed omissions are attributed to the responsible builder, who forfeits the block reward -- making compliant disclosure the profit-maximizing strategy.

The same architectural pattern -- delegating a fairness-critical decision to a committee under a 1-of-$n$ honest assumption -- underlies a parallel line of research on MEV mitigation through data-independent Order Policy Enforcement (OPE), either via fair-ordering protocols~\cite{kelkar2020order} or via content-oblivious ordering.
Wadhwa et al.~\cite{wadhwa2024ope} prove that this assumption is fundamentally fragile: no protocol within a broad framework capturing the above OPE schemes can enforce its ordering policy when all committee members are rational and an MEV-extracting permutation of strictly higher utility exists.

\vspace{1mm}
\paragraph*{On Observation Window Selection}
\label{discussion:observation_window}
The choice of observation window reflects a trade-off between classification reliability and residual misclassification risk.
Longer windows evaluate proposer behavior over more block-construction decisions, but also raise the risk that early (misclassified) non-altruistic behavior determines the classification for the entire window.
A proposer that relied on MEV-Boost in early periods, but later acted altruistically, would still be excluded if the observation window is chosen too long -- and a single misclassification propagates to more proposers through our cluster-level exclusion.
Shorter windows, by comparison, base the classification of many proposers on only a single block, limiting reliability.
To assess whether our findings are sensitive to this trade-off, we conducted five complementary measurements: Q1, Q2, Q3, Q4, and the aggregated period (2025), with quarterly results reported in \Cref{tab:proposers}.
The individual quarters serve as short-window references, while in the aggregated period the median proposer's classification is based on more than one block.
Q1 yields a potentially altruistic share of \SI{2.12}{\percent}.
This share stabilizes within \SI{1.64}{\percent}--\SI{1.68}{\percent} across the quarters Q2, Q3, and Q4, close to the aggregated period (\SI{1.55}{\percent}), indicating that our upper bound is robust to the choice of observation window.
The lower bound, in contrast, exhibits a stronger window dependence: all individual quarters yield higher shares (\SI{0.52}{\percent}--\SI{0.79}{\percent}) than the aggregated window (\SI{0.41}{\percent}).
While a lower aggregated share follows by construction, the pronounced gap points to a possible explanation: the per-block criteria of Classifiers~\textup{\Circled{\RNum{5}}} and~\textup{\Circled{\RNum{6}}} may be too strict, such that even genuinely altruistic proposers satisfy them in some blocks but not in all (cf.\ \Cref{sec:results}).
Under this reading, the aggregated window yields the most conservative lower bound, and the interval $[\SI{0.41}{\percent};\SI{1.55}{\percent}]$ provides a meaningful estimate of the genuine share of altruistic proposers.

\vspace{1mm}
\paragraph*{Transaction Censorship Beyond MEV}
Transaction censorship may be motivated by MEV or by external compliance requirements; both conflict with Ethereum's censorship-resistance ideal~\cite{buterin2021design}.
Although censorship would thus be a natural indicator, we do not use censorship as a classification criterion.
First, Ethereum mempools are not mutually consistent, so the absence of a transaction from a block may reflect propagation effects rather than intentional exclusion and thus provides no reliable evidence of non-altruism.
Second, state-driven censorship is too rare in our observation window to support meaningful inference: the only restriction known to us, the OFAC SDN list, was lifted for Tornado Cash -- the only actively used sanctioned service -- during the period~\cite{tornado_cash_ofac}, and interactions with sanctioned addresses were rare even before.
Among potentially altruistic proposers, we observed no mempool transactions involving sanctioned addresses, so we cannot assess compliance with state-imposed censorship.
Proposers willing to censor such transactions may thus remain undetected; our estimate could therefore be slightly inflated.

\vspace{1mm}
\paragraph*{Validity under ePBS}
With Enshrined Proposer-Builder Separation~\cite{EIP7732}, the outsourcing of block construction to builders becomes a protocol-native feature.
Expected to activate with Ethereum's Glamsterdam upgrade~\cite{eip7773} in the second half of 2026, ePBS eliminates the need for MEV-Boost relays as trusted intermediaries.
This removes the basis for classifying such behavior as a protocol deviation -- and thus as a proxy for non-altruism.
The underlying incentive structure, however, remains largely unchanged.
Proposers remain unable to influence block contents prior to propagation, and censorship decisions continue to reside with builders.
Likewise, proposers still lack any mechanism to prevent unfair block construction.
Our findings therefore remain directly relevant, as proposer behavior is still governed by the same rational incentives as before.
Proposers who previously chose to outsource block construction -- thereby revealing a preference for revenue-maximizing strategies -- appear unlikely to reverse this choice once outsourcing is enshrined.
More importantly, this revealed preference carries over to fairness mechanisms.
Proposers who already exploit protocol-granted freedoms for revenue can be expected to do the same within a fairness mechanism -- for instance, by accepting bribes for deviating from its intended operation.

\vspace{1mm}
\paragraph*{Reproducibility and Ethics}
All data can be reconstructed from the sources listed in \Cref{sec:methodology}; XOF transactions can be identified using public mempool archives.
Analysis scripts are publicly available~\cite{EthereumAltruisticProposers2026} to support reproducibility.
The study operates on pseudonymous addresses; we deliberately refrain from de-anonymization for ethical reasons.
This choice comes at a cost: clusters we treat as distinct may overlap, in which case our method would overestimate the altruistic share.
Nevertheless, we expect the resulting bias to be minor.

\section{Conclusion}
\label{sec:conclusion}
By progressively excluding proposers for whom we observe evidence of non-altruistic behavior, we empirically estimate the share of Ethereum proposers exhibiting altruistic behavior over the full year 2025.
Buterin's statement that ``at least a few percent of proposers are altruistic''~\cite{buterin2021design} is, depending on interpretation, only partially supported by our analysis.
By combining an upper bound that excludes proposers with clear non-altruistic behavior with a strict lower bound restricted to proposers for whom altruistic behavior is undisputed, we show that the share of genuinely altruistic proposers falls within the interval $[\SI{0.41}{\percent};\SI{1.55}{\percent}]$.
Both bounds remain reasonably stable across the individual quarters of 2025.
This range is nearly an order of magnitude below the share that current committee-based fairness mechanisms would require to provide meaningful censorship resistance.
Achieving a sufficiently high probability of containing at least one altruistic member requires committees of at least \num{128} proposers -- an order of magnitude larger than the \num{16}-member committees currently discussed.
At this scale, the influence of any single member on the resulting block diminishes substantially.
The viability of specific fairness mechanisms therefore warrants further investigation, particularly with respect to their incentive structure and the extent to which individual members can benefit from deviation.
Mechanisms relying solely on participants' altruism appear insufficient under the conditions we observe.

\begin{acks}
We gratefully acknowledge the Decentralized Systems Lab at Yale University for providing the research data used in this study.
This work was supported by the \href{https://kastel-labs.de/}{KASTEL Security Research Labs}.
\end{acks}

\begingroup
\makeatletter
\let\prism@lbibitem\@lbibitem
\let\prism@bibitem\@bibitem
\def\@lbibitem[#1]#2{\filbreak\prism@lbibitem[#1]{#2}\interlinepenalty=10000\clubpenalty=10000\widowpenalty=10000\brokenpenalty=10000}
\def\@bibitem#1{\filbreak\prism@bibitem{#1}\interlinepenalty=10000\clubpenalty=10000\widowpenalty=10000\brokenpenalty=10000}
\makeatother
\interlinepenalty=10000
\clubpenalty=10000
\widowpenalty=10000
\brokenpenalty=10000
\bibliographystyle{ACM-Reference-Format}
\bibliography{sample-base}
\endgroup

\end{document}